\documentclass[twocolumn,showpacs]{revtex4}

\usepackage{graphicx}

\newcommand\be{\begin{equation}}
\newcommand\ee{\end{equation}}

\newcommand\ba{\begin{eqnarray}}
\newcommand\ea{\end{eqnarray}}

\begin{document}

\title{Dynamic Nonlinear X-waves for Femtosecond Pulse Propagation in Water}

\author{M. Kolesik}

\author{E. M. Wright}

\author{J. V. Moloney}

\affiliation{ACMS and Optical Sciences Center, University of
Arizona, Tucson, AZ 85721}

\begin{abstract}
Recent experiments on femtosecond pulses in water displayed long
distance propagation analogous to that reported in air. We verify
this phenomena numerically and show that the propagation is
dynamic as opposed to self-guided. Furthermore, we demonstrate
that the propagation can be interpreted as due to dynamic
nonlinear X-waves whose robustness and role in long distance
propagation is shown to follow from the interplay between
nonlinearity and chromatic dispersion.
\end{abstract}


\maketitle

The nonlinear Schr\"odinger equation (NLSE) in two or more
dimensions is ubiquitous in physics as a model for weakly
interacting nonlinear and dispersive waves, and arises in such
diverse areas as Langmuir waves in plasmas, weakly interacting
Bose-Einstein condensates, and optical propagation in nonlinear
dielectrics \cite{DyaNewPus92}. The ubiquity of the NLSE means
that new solutions or paradigms that arise in one area can extend
into other areas. For example, previous experiments have shown
that femtosecond (fs) pulses can propagate long distances though
air while maintaining an almost constant fluence profile
\cite{BraKorLiu95,NibCurGri96,BroChiIlk97}. (Here long distance
means that filaments of wavelength $\lambda$ and radius $r_0$
persist for distances much longer than their associated Rayleigh
range $\pi r_0^2/\lambda$.) Although these results initially
suggested a self-guiding mechanism, with self-focusing balanced by
plasma defocusing, numerical simulations revealed that the
propagation is highly dynamic, and this led us to the paradigm of
dynamic spatial replenishment, whereby the propagating pulse
collapses, the collapse is arrested, and the process is repeated
several time as the collapse is replenished from spatially
delocalized power \cite{MleWriMol98}. One would then expect
analogous phenomena in other fields, and indeed the dynamic
spatial replenishment model in air has analogies with the
Bose-Nova phenomenon in atomic gases \cite{DonClaCor01}.

Here our goal is to elucidate the physics underlying recent
observations \cite{DubTamDio03} of long distance propagation of fs
pulses in water. Long distance propagation has previously been
explored in glass \cite{TzoSudFra01} and there are clear
differences with respect to air propagation. For example, in
silica glass, and water also, normal group-velocity dispersion
(NGVD) plays a much more dominant role in comparison to air, and
this gives rise to nonlinear pulse-splitting
\cite{ZhaLitPet86,ChePet92,Rot92,RanSchGae96,Fibich2003}. Using numerical
simulations we first verify the reported properties for long
distance propagation in water, and we then perform diagnostic
simulations to elucidate the underlying physics. In particular, we
show that long distance propagation in water is given a natural
explanation by combining the paradigms of nonlinear
pulse-splitting and nonlinear X-waves. Nonlinear X-waves arise
from the combination of diffraction, NGVD, and self-focusing, and
have recently been introduced and examined theoretically
\cite{ConTriDit03} and experimentally \cite{DitValPis03}. Our main
conclusion is that long distance propagation in water is best
understood in terms of nonlinear X-waves.

Our model for fs pulse propagation in water is based on the
propagation equation for the spectral amplitudes of the
Bessel-beam expansion of the axially symmetric electric field:
\begin{equation}
{\partial E(\omega,k,z)\over \partial z} = i
\sqrt{{\omega^2\epsilon(\omega)\over c^2} - k^2} E(\omega,k,z) +
\frac{i \omega }{2 c n_b} P(\omega,k,z) . \label{eq:propag}
\end{equation}
This is a scalar version of the Unidirectional Pulse Propagation
Equation solved in the $z$-direction~\cite{Kolesik2002}, and in
the paraxial approximation becomes equivalent to the Nonlinear
Envelope Equation of Brabec and Krausz~\cite{Brabec97}. We utilize
a tabulated representation~\cite{Kolesik2003b} of the complex
frequency-dependent water permitivity $\epsilon(\omega)$. In
Eq.~(\ref{eq:propag}), $k$ stands for the {\em transverse}
wavenumber of each Bessel-beam component, and the nonlinear
polarization $P(\vec{r},t)=\Delta\chi(\vec{r},t)E(\vec{r},t)$ is
calculated in the real-space representation from the local
nonlinear modification of the material susceptibility:
\begin{equation}
\Delta\chi(\vec{r},t) =   2 n_b n_2 I + \chi_{\rm pl}( \rho ) + i
{n_b^2\beta^{(K)}\over k_0} I^{K-1}
\end{equation}
The first term represents the instantaneous optical Kerr effect
with nonlinear coefficient $n_2 = 2.7\times 10^{-20}$ m$^2$/W, the
second is the free-electron induced susceptibility change
$\chi_{\rm pl}(\rho)=(i\rho e^2/\epsilon_0 m_e\omega_0)/(1/\tau_{\rm c} - i \omega_0)$,
and the
third term represents multi-photon ionization (MPI) energy losses,
with MPI coefficient $\beta^{(K)}$. The evolution of the
free-electron density is described by the equation
\begin{equation}
{\partial\rho\over\partial t} = {\sigma\over n_b^2 E_g}\rho I +
{\beta^{(K)}\over K \hbar \omega_0} I^K - a \rho^2 \ ,
\end{equation}
where the three terms describe avalanche ionization
with $\sigma = (e^2\tau_{\rm c} n_b/m_e \epsilon_0 c)/(1 + \omega_0^2\tau^2_{\rm c})$ 
the cross-section for inverse bremsstrahlung, electron
generation via MPI, and electron-ion recombination, respectively.
The plasma model is parameterized by the collision time $\tau_c =
10^{-14}$s and recombination rate $a=2\times 10^{-15}$ m$^3$/s
\cite{Docchio88}. The MPI rate is calculated from the formula
given in~\cite{Kennedy95b,Kennedy95a} for $E_g=7$ eV.

The simulation parameters were chosen to match the experiment of
Ref.~\cite{DubTamDio03} as closely as possible for an incident
pulse of center wavelength $\lambda=527$ nm. The initial pulse
amplitude within the water sample was chosen as a focused Gaussian
of spot size $w_0=99$ $\mu$m, pulse duration $\tau_p=170$ fs, an
initial radius of curvature corresponding to a lens of focal
length $f=5$ cm, and we varied the pulse input energy between
$0.5$ $\mu$J and $2.5$ $\mu$J. We note that, as in the experiment,
the choice of initial focusing and spot size are essential for
realizing long-distance propagation.

\begin{figure}[t]
\centerline{\scalebox{0.55}{\includegraphics{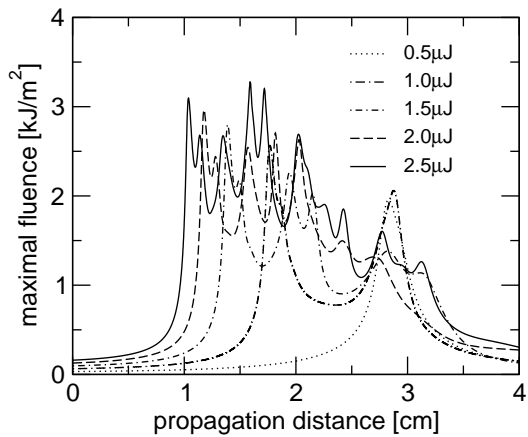}}} \caption{
Maximum fluence over the transverse plane 
for several initial pulse energies.}
\label{fig:fluence}
\end{figure}

First we demonstrate that our simulations reproduce the basic
features of the experiment in Ref.~\cite{DubTamDio03}.
Figure~\ref{fig:fluence} shows the maximum fluence
(time-integrated intensity) over the transverse plane versus
propagation distance $z$ in the water cell for several input pulse
energies. An elevated and sustained fluence is taken as a sign
that a light filament has formed. Similar to the experiment, at an
input pulse energy $0.5$ $\mu$J a filament just starts to form but
is terminated by diffractive spreading shortly after. At higher
energies, we obtain filaments whose fluence fluctuates with $z$
but which persist over the scale of a couple of centimeters in
quantitative agreement with the experiment.
\begin{figure}[t]
\centerline{\scalebox{0.55}{\includegraphics{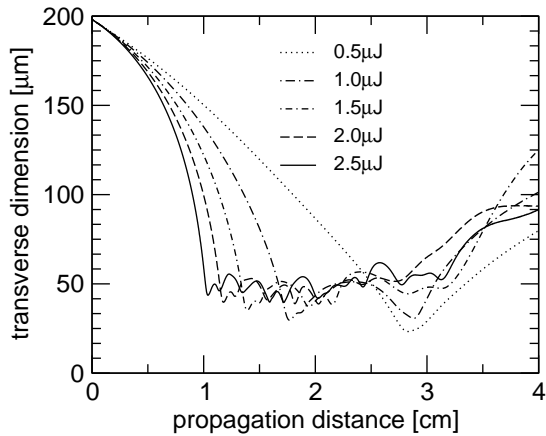}}} \caption{
Transverse dimension of the fluence profile as a function of
the propagating distance showing that a filament of $\simeq 50$
$\mu$m diameter persists over a distance of around 2~cm.}
\label{fig:dimension}
\end{figure}
Furthermore, the transverse fluence profiles at different
propagation distances reveal an almost constant filament diameter
of around $\simeq 50$ $\mu$m over the two centimeter range of the
filament. This is shown in Fig.~\ref{fig:dimension} where the
$1/e^2$ diameter, obtained by approximating the spatial integral
of the fluence profile with the on-axis fluence times 
$\pi d^2 /8$ (corresponding to a gaussian),
is shown versus propagation distance for several pulse
energies. Thus, our simulation model agrees with the experiment of
Ref. \cite{DubTamDio03} in which the filaments also persist for
around two centimeters and have diameters $\simeq 60$ $\mu$m.

\begin{figure}[th]
\centerline{\scalebox{0.5}{\includegraphics{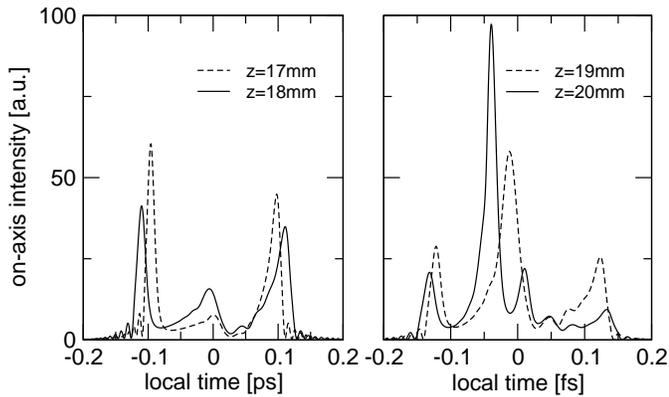}}}
\caption{On-axis intensity as a function of local time for an
input pulse energy $1.5$ $\mu$J and a variety of propagation
distances.} \label{fig:pulseG}
\end{figure}

The fluence maxima versus propagation distance in
Fig.~\ref{fig:fluence} suggest that multiple re-focusings occur.
However, in contrast to propagation in air where the collapse is
arrested dominantly by plasma defocusing \cite{Cou03}, in water,
and also in glass \cite{RanSchGae96}, NGVD is a key player. To
diagnose this we have performed comparative simulations
corresponding to Figs.~\ref{fig:fluence} and \ref{fig:dimension}
but with the plasma turned off, and the results (not shown),
though clearly changed, are very similar in terms of the predicted
2~cm length scale for the filament and $\simeq 50$ $\mu$m filament
diameter. This indicates that the plasma is not essential to
understanding this propagation regime: It somewhat slows down the
dynamics and results in slightly thicker filament, because it
helps to arrest the collapse, but the main collapse arresting
mechanism is chromatic dispersion. It has previously been shown
that a signature of collapse arrested by NGVD is nonlinear pulse
splitting in the time domain
\cite{ZhaLitPet86,ChePet92,Rot92,RanSchGae96}.
Figure~\ref{fig:pulseG} shows the results of our simulations,
including the plasma, for the on-axis intensity as a function of
local time for an input pulse energy $1.5$ $\mu$J and a variety of
propagation distances after the filament first appears. The
left-hand plot shows the pulse after the first pulse splitting,
and the split daughter pulses are seen to move apart with
distance. This figure is in keeping with the usual picture of
pulse splitting \cite{ZhaLitPet86,ChePet92,Rot92,RanSchGae96}, and
the evident asymmetries derive from inclusion of the plasma and
the fact that we capture the spatio-temporal focusing terms to all
orders and retain the full dispersion landscape as opposed to
keeping only second-order GVD. We remark that the daughter pulses
do not undergo subsequent cascade splittings \cite{ZhaLitPet86},
but rather after the first splitting, energy is replenished into
the center of the local time domain around $\tau=0$ which then
grows, and it is this replenished center pulse that is subject
to further pulse splitting. This is shown in the right-hand plot
in Fig.~\ref{fig:pulseG}. For larger propagation distances this
process of pulse splitting, temporal replenishment of the on-axis
pulse, followed by splitting of the new center pulse, can repeat
itself several times, and this gives rise to the multi-peaked
structures in Fig.~\ref{fig:fluence}, and the illusion that the
filament is propagating in a self-guided manner. For comparative
purposes we have repeated these simulations with the plasma turned
off (results not shown) and we find that the same basic pulse
splitting coupled with temporal replenishment picture emerges for
the pulse propagation. The main difference is that the on-axis
intensity profiles are more symmetric, have more structure and
show evidence of shock phenomena since the smoothing action of the
plasma is absent. This tells us that the propagation is dominated
by an interplay between dispersion and nonlinearity, in which the
delayed nonlinear responses such as plasma do not play a crucial
role.

Since the multiple pulse splittings are distinct from the cascade
splitting predicted for a medium exhibiting NGVD, this begs the
question as to what physics produces the temporal replenishment of
the on-axis pulse? Our proposal here for the physical mechanism
that produces the temporal replenishment is that it is due to
dynamic nonlinear X-waves, and we shall make our case below.
Nonlinear electromagnetic X-waves have only very recently been
introduced theoretically \cite{ConTriDit03} and produced
experimentally \cite{DitValPis03} in the field on nonlinear
optics. They are stationary (z-invariant) nonlinear solutions that
result from the combination of linear diffraction, second-order
NGVD, and nonlinear self-focusing. These X-waves propagate along
$z$ with unchanging intensity profile, and if one plots a
space-time (local time) slice the resulting intensity profile has
a central peak and arms that form a characteristic X shape (see
Fig. 1 of Ref. \cite{ConTriDit03}). Likewise, if one calculates
the space-time Fourier-transform of a stationary X-wave it's
intensity spectrum in $(k,\omega)$-space, $k$ being the magnitude
of the transverse wave vector and $\omega$ the frequency, likewise
has an X-structure (see Fig. 3 of Ref. \cite{ConTriDit03}). It has
been shown that the X-shape in $(k,\omega)$-space should follow
the linear dispersion characteristics since they determine the
relation between the propagation angle and frequency for
phase-matched off-axis conical emission
\cite{ConTriDit03,LutNewMol94}.

For the series of simulations presented here X-shaped textures can
be discerned in the intensity profiles in both the space-time and
$(k,\omega)$ domains past the first pulse splitting. The X-wave
texture appears in the space-time domain in the region between the
daughter pulses after splitting (not shown). This texture reflects
optical energy that is transported from the off-axis regions and
which serves as a reservoir for the replenishment of the pulse
center as illustrated in Fig.~\ref{fig:pulseG}.
\begin{figure}[t]
\centerline{\scalebox{0.72}{\includegraphics{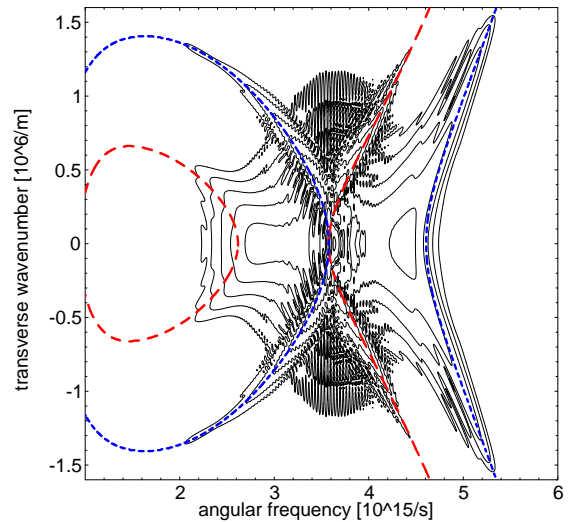}}}
\caption{ (color online) Contour plot of the logarithmic spectral intensity in
$(k,\omega)$-space for $z=1.7$~cm after the first pulse splitting.
} \label{fig:Xspc}
\end{figure}
The X-wave signature for our pulse dynamic is much
more evident in $(k,\omega)$-space. This is shown in
Fig.~\ref{fig:Xspc} which shows a representative example of a
contour plot of the pulse spectrum at $z=1.7$ cm (fully including
plasma effects). The fine, ``interference'' structure in this
figure results from the superposition of the spectra from both
split daughter pulses. For longer propagation distances, after
multiple pulse splittings the intensity spectra show similar
features but become more complicated. The central X-shaped region
where spectral energy concentration occurs is clearly visible.
However, one quickly realizes that the X-shape has a slightly
different angle compared to that expected for pure X-waves.
Nevertheless, as we show next not only the spectrum in
Fig.~\ref{fig:Xspc} can be explained using the paradigm of X-waves, but in
the process we also gain further insight into the role of
nonlinearity in X-wave propagation.

To proceed we make use of the three-wave mixing picture we
recently employed to provide a qualitative explanation of
supercontinuum generation in bulk media~\cite{Kolesik2003a}. The
nonlinear light-matter interaction creates a material response
reflected in the change in the on-axis material susceptibility
$\Delta\chi\approx\Delta\chi(t-z/v_g,\tau_{\rm slow})$ that
propagates predominantly as a $z$-invariant shape modulated on a
relatively slower time scale $\tau_{\rm slow}$, that we hereafter
neglect. This response usually exhibits multiple peaks (roughly
corresponding to multiple intensity maxima), which propagate with
slightly different group velocities $v_g$. New optical frequencies
are generated through the scattering of incident fields of these
``material waves'' to produce a third wave. An incident optical
wave at $(\omega,\vec{k}(\omega))$ will predominantly scatter into
the wave component at $(\Omega,\vec{s}(\Omega))$ that satisfies
the phase matching condition
\begin{eqnarray}
\sqrt{\frac{\Omega^2\epsilon(\Omega)}{c^2} - {s}_\perp^2} &=&
{(\Omega - \omega)\over v_g} + \sqrt{
\frac{\omega^2\epsilon(\omega)}{c^2}- k_\perp^2} , \cr {s}_\perp
&=& k_\perp + m_\perp , \label{eq:phasematch}
\end{eqnarray}
where $m_\perp$ is the transverse Fourier component of the
material wave. As explained in Ref.~\cite{Kolesik2003a}, this
phase matching condition is not strictly enforced, but
nevertheless, it is useful in identifying the loci in the spectral
domain where the new spectral components are predominantly
generated.

In the present case, despite strong white-light generation most of
the spectral energy is concentrated in the vicinity of the
original carrier frequency $\omega_0$ around zero transverse
wavenumber. Therefore, we set
$(\omega,\vec{k}(\omega))=(\omega_0,k_0)$ for the incident optical
field in Eq.~(\ref{eq:phasematch}) together with $m_\perp = 0$ to
find the loci $(\Omega,\vec{s}(\Omega)$ where the the three-wave
mixing is near phase-matched and most effective. To solve
(\ref{eq:phasematch}) we need to estimate the group velocity $v_g$
for the strongest response peaks. We have obtained these estimates
from the raw data for the on-axis value of the material response
$\Delta\chi(\vec{r},t)$ by measuring the shift of the response
peaks for both split daughter pulse with propagation distance. As
a result we obtained {\it two} solutions to
Eq.~(\ref{eq:phasematch}) corresponding to the scattering on the
two strongest response peaks. The loci $(\Omega,\vec{s}(\Omega)$
where the the three-wave mixing is most effective according to the
above prescription are shown as dashed-lines in
Fig.~\ref{fig:Xspc}, the short-dashed line represents scattering
from the material response generated by the leading daughter
pulse, while the long-dashed line is due to trailing edge
daughter pulse, and the phase-matching curves clearly coincide
with the strong features of the spectrum. Thus, our phase matching
argument captures the central X shape perfectly, including the
curvature of its arms (due to non-paraxial effects), 
and it
furthermore explains the occurrence of the low- and high-frequency
``ridges'' in the spectrum. An important point is that each half
of the X is actually generated by one daughter pulse. If we used
the group velocity $v_g(\omega_0)$ taken at the central wavelength
instead of the actual $v_g$'s, we would obtain a single X-shaped
locus expected for ``normal'' X-waves. This turns out to have the
same center, and a small difference in the angles between the arms
are due to $v_g \neq v_g(\omega_0)$. Thus, even after multiple
pulse splitting, the split-off pulses ``cooperate'' and contribute
to the intensity spectrum in the same X-shaped region in the
$(\omega,\vec k)$ space. This is the origin of the universality
and robustness of the X-waves, namely, irrespective of the details
of the temporal dynamics, they tend to concentrate their spectral
energy around the locus that supports the $z$-invariant
propagating waves. Though they may not form z-invariant light
bullets, the X-wave character therefore clearly manifests itself
in long-distance propagation in water. Finally, we note that the
above argument doesn't depend on details of the nonlinearity or of
the linear dispersion. Thus, dynamic X-waves should be inherent to
many systems, though whether their signature can be observed
depends on the relative magnitudes of other competing effects and
initial pulse conditions.

In summary, we have provided diagnostic numerical simulations to
show that the recent experimental observations long distance
propagation in water \cite{DubTamDio03} may be interpreted by
combining the paradigms of both nonlinear pulse splitting and
dynamic X-waves that develop from the split pulses and replenish
the pulse center. Our results therefore reveals intimate
connections between long distance propagation in condensed matter
and nonlinear electromagnetic X-waves.

We thank M. Mlejnek (Corning) and Q. Feng for discussions of pulse
propagation in water over several years. This work is sponsored by
the U.S. Air Force Office of Scientific Research (AFOSR), under
grants AFOSR-F49620-00-1-0312 and AFOSR-F49620-03-1-0194.

\end{document}